\documentclass[aps,prb,twocolumn,showpacs]{revtex4}

\usepackage{graphicx}
\usepackage{dcolumn}
\usepackage{amssymb}
\usepackage{amsmath}

\begin{document}

\title{Effect of stress-triaxiality on void growth in dynamic 
fracture of metals: a molecular dynamics study}

\author{E.\ T.\ Sepp\"al\"a}
\author{J.\ Belak}
\author{R.\ E.\ Rudd}

\affiliation{Lawrence Livermore National Laboratory, Condensed Matter Physics 
Division, L-415, Livermore, CA 94551, USA}
\date{\today}

\begin{abstract}
The effect of stress-triaxiality on growth of a void in a three
dimensional single-crystal face-centered-cubic (FCC) lattice has been
studied. Molecular dynamics (MD) simulations using an embedded-atom
(EAM) potential for copper have been performed at room temperature and
using strain controlling with high strain rates ranging from
$10^7$/sec to $10^{10}$/sec. Strain-rates of these magnitudes can be
studied experimentally, {\it e.g.} using shock waves induced by laser
ablation.  Void growth has been simulated in three different
conditions, namely uniaxial, biaxial, and triaxial expansion.  The
response of the system in the three cases have been compared in terms
of the void growth rate, the detailed void shape evolution, and the
stress-strain behavior including the development of plastic
strain. Also macroscopic observables as plastic work and porosity have
been computed from the atomistic level.  The stress thresholds for
void growth are found to be comparable with spall strength values
determined by dynamic fracture experiments. The conventional
macroscopic assumption that the mean plastic strain results from the
growth of the void is validated.  The evolution of the system in the
uniaxial case is found to exhibit four different regimes: elastic
expansion; plastic yielding, when the mean stress is nearly constant,
but the stress-triaxiality increases rapidly together with exponential
growth of the void; saturation of the stress-triaxiality; and finally
the failure.
\end{abstract}


\maketitle

\section{Introduction}
\label{intro}

Ductile fracture of metals commonly occurs through the nucleation,
growth and coalescence of microscopic voids.~\cite{knott} Much can be
learned about the ductile fracture through the study of these voids.
A particularly interesting case is the dynamic fracture of ductile
metals,~\cite{DynFract,Curran,McClintock,Barbee} in which the strain
rates are so high that processes such as diffusion operating on
relatively long time scales may be neglected, while inertial effects
become relatively important.  Void growth is driven by the need to
relax tensile stress that builds up in the system, and to minimize the
associated elastic energy.  The material around a void deforms
plastically in order to accommodate the void growth.  Naturally, the
plastic deformation results from a local shear stress, which may arise
from the applied stress, but it also may arise from the stress field
of the void even if the applied stress is hydrostatic.  So the
expectation is that the evolution of the plastic zone, and hence the
growth of the void, is influenced by the degree of stress triaxiality;
{\it i.e.}, the ratio of the mean (hydrostatic) stress to the shear
stress.  It is this relationship that we study here by varying the
triaxiality of the loading.  In particular, we conduct simulations in
which one, two or three directions of the system are expanded,
producing a state of uniaxial, biaxial or triaxial strain,
respectively.  Variation in the triaxiality of the strain causes
variation in the triaxiality of the stress state, where it should
be noted that uniaxial (biaxial) strain does not imply pure uniaxial
(biaxial) stress.

Besides dynamic crack propagation experiments, dynamic fracture can be
measured for instance in shock physics or spallation experiments, to
which the simulations performed here are compared.  Various techniques
are employed to generate the shock waves: Hopkinson bar, gas gun, high
explosives, and laser ablation. With Hopkinson bar the strain-rates
$\dot{\varepsilon}$ usually are of the order $10^2 - 10^4$/sec, in gas
gun of $10^5$/sec, with high-explosives even higher strain-rates can
be produced, and with lasers strain-rates exceeding $10^7$/sec are
attained. In a gas gun for instance the fracture results from
essentially one-dimensional shock loading.  Two compressive shock
waves are generated by the impact of a flier on a metal target,
propagate away from each other, reflect from opposite free surfaces
becoming tensile release waves and finally come into coincidence
again.  If the combined tensile stress exceeds the rupture strength of
the material, the metal fails, after some incubation time, producing a
fracture surface.  In strong shocks, a scab of material may spall from
the back side of the target and fly off.  Spallation
experiments~\cite{Kanel} for single and polycrystal copper report
spall strength values of $\sigma^* \simeq 3-4$~GPa at strain-rates
$\dot{\varepsilon} \simeq 2-3\times 10^5$/sec, and scaling between the
spall strength and strain-rate, $\sigma^* \sim
\dot{\varepsilon}^{0.2}$.

In this study of dynamic fracture in ductile metals at high
strain-rates ($10^7$ -- $10^{10}$/sec) we have concentrated on void
growth starting from a single crystal copper lattice containing an
infinitely weakly bound inclusion or a pre-existing nanoscale void.  The
lattice is initially free of other defects.  We have focused on the
effect of stress-triaxiality on void growth. In some fracture
experiments, for example in necking and cup-cone
fracture,~\cite{tvergaard84} the uniaxial strain produces a stress
state that transitions rapidly to triaxial state due to the plastic
flow during the course of loading.  It is during
the triaxial phase that void growth and failure take place.  Because
of the connection with shock experiments, the stress triaxiality study
done here is carried out using strain control, and it is the strain
that is varied from uniaxial to biaxial to triaxial.

Much of the damage modeling of metals has been carried out at
mean-field or continuum level based on constitutive theories. The
continuum models concentrate especially on two areas: macroscopic
crack growth phenomenon~\cite{Tvergaard93,tvergaard98} and studies of
porosity, {\it i.e.} behavior of an array of voids, at sub-grain level
during
loading.~\cite{McClintock2,Rice,Green,Needleman,Brown,Shima,Gurson,Tvergaard,Cocks,Tvergaard2,Huang,Cortes,tvergaard95,tvergaard952,Giessen,Shu,Wu}
In the latter area for example the locus of yield surfaces in stress
space has been studied, which is related to the question of the effect
of stress-triaxiality studied here. Of the void growth studies
especially the Gurson model~\cite{Gurson} is commonly used to model
cavitation (the development of  porosity) at the sub-grid level 
in what is termed as damage modeling. 
These continuum calculations often
assume that the matrix material, where the voids are embedded, is
elastically rigid and plastically incompressible, and the dilation of
the void-matrix aggregate is completely due to the void growth. Of
particular interest, and relevance in terms of this study, is a single 
crystal plasticity study of void growth.~\cite{mchugh}  The
calculations are typically done by determining approximate solutions for
integrals of incremental equations of virtual work using the finite
element method. Continuum modeling has been used to study some of the
phenomena addressed in this paper such as the effect of triaxiality on
void growth and void shape changes.~\cite{castaneda} The validity of
the approximate solutions of incrementals limits the strain-rates to
be rather low compared to the strain-rates used in this study.

In order to characterize the void growth not only with macroscopic
quantities and at the continuum level, but to investigate what happens
at the atomistic level, we have employed molecular dynamics (MD)
simulations. MD simulations enable us to see what the effects are on
the void surface at the single atom level, when it grows while the
total system yields.  This Article presents work, which is an
extension to the work done earlier by some of this Article's authors
of void growth in a single crystal copper with hydrostatic loading,
void nucleation and growth in single and polycrystalline
copper.~\cite{belak,belak2,belak3,rudd,ices,disl} Molecular dynamics
simulations of void growth in single crystal copper have also been
conducted by other groups for slab geometries of interest in the
semiconductor chip metalization problem.~\cite{farrisey,gungor} The
effectively two-dimensional, thin film systems are in contrast to
three dimensional bulk systems studied here. In some cases MD 
simulations of plasticity and crack propagation have been as large 
as billion atoms.~\cite{fabraham}

This Article is organized as follows. It starts in
Section~\ref{method} with an overview of the MD method used and the
simulations which have been carried out in this study.  Exploration of
the results of the simulations starts in Section~\ref{stressstrain} by
the study of the mean or hydrostatic stress versus strain as well as
the deviatoric part of the stress tensor, von Mises stress, which is used
to measure the shear stress, and stress-triaxiality versus strain for
all the simulated strain-rates and modes of expansion.
Section~\ref{plasticstrain} concentrates on the macroscopic plastic
quantities, such as mean and equivalent plastic strain, plastic work
and its relation to the temperature.  The evolution of the void in
terms of its growth and shape changes is studied in
Section~\ref{voidevol}. Section~\ref{summary} summarizes the results
and compares different measured quantities with each other
concentrating on one of the simulations, uniaxial strain with
strain-rate $10^8$/sec. The Article is concluded with discussions of
the results and suggestions for future studies in
Section~\ref{discussion}.

\section{Method and Simulations}
\label{method}

\subsection{Strain-controlled Molecular Dynamics}
\label{MDmethod}

In this atomistic-level study of void growth, the simulations have
been done using empirical embedded-atom (EAM) potentials in classical
molecular dynamics~\cite{allen} following the scheme developed
earlier.~\cite{belak,belak2,belak3} The copper EAM potential we have
used is due to Oh and Johnson.~\cite{oh1,oh2}

The system, in which the simulations are done, is a three-dimensional
single-crystal face-centered-cubic (FCC) lattice in a cubic box with
$\{100\}$ faces.  Periodic boundary conditions are used in all the
three directions so that there are no free boundaries in the system
apart from the void. Equivalently the system can be imagined to
consist of an infinite periodic array of voids.  Note that periodic
boundary conditions have also been used in continuum models of void
growth, but in the continuum modeling of void growth in isotropic 
materials the calculations are done in a
reduced cell, which exhibits one quarter of the box in two dimensions,
and one eighth in three dimensions, and the behavior of other areas
are derived from the symmetries.  We use the full cubic box because
the cubic symmetry present in the continuum is broken in MD at finite
temperature, and processes such as dislocation nucleation at the void
surface would be over-constrained in a reduced simulation box.

In the simulations, the system is brought to thermal equilibrium at
room temperature, $T=300$~K, with a commonly used
thermostat~\cite{hoover} and at ambient pressure, $P \simeq 0$~MPa,
keeping the volume constant. After that a spherical void is cut in the
middle of the system, later the thermostat is turned off, and the
dilational strain is applied uniformly with a constant strain-rate
$\dot{\varepsilon}$.  The removal of the atoms in the spherical region
may be considered to simulate the instantaneous separation of the
matrix material from an infinitely weakly bound inclusion.  The
uniform expansion in these strain-controlled simulations is applied
through rescaling the coordinates as in the Parrinello-Rahman
method.~\cite{parrinello} Technically the three Cartesian coordinates
of the atoms are rescaled to the unit-box, each coordinate $S_\alpha
\in [0,1)$.  When calculating the forces and velocities, as well as
updating the new positions of the atoms, the unit-box is multiplied by
a diagonal scaling matrix ${\mathcal H} = \{l_x,l_y,l_z\}$, where
$l$'s are the simulation box's side lengths, to compute the true
positions of the atoms,
\begin{equation}
{\bf x} = {\mathcal H} {\bf S}.
\label{eq_scale}
\end{equation}
This scaling matrix ${\mathcal H}$ is updated each time-step, when the
load is applied, by multiplying the initial matrix ${\mathcal H}_0$
with the sum of the unit matrix and the strain matrix ${\mathcal E} =
t\dot{\mathcal E}$,
\begin{equation}
{\mathcal H}(t) = {\mathcal H}_0  ({\mathbb I} + t \dot{\mathcal E}).
\label{eq_update}
\end{equation}
For our purposes the strain-rate matrix $\dot{\mathcal E}$ is always
diagonal, since neither rotation nor simple shear type strains are
studied. In the triaxial case all the terms in the diagonal are equal;
in the uniaxial there is a single non-zero term; and in the biaxial
case two of the three diagonal terms differ from zero and are
equal. Prior to expansion the system is cubic, its scaling matrix
${\mathcal H}_0$ is diagonal, and all the terms are equal and
correspond to the equilibrium size at ambient pressure.  Hence the
scaling matrix ${\mathcal H}$ remains diagonal throughout the
simulation, and the strain in each case is in a $\langle 100 \rangle$
direction.

In fracture and plasticity simulations the first quantity to consider
is the stress-strain behavior. With the strain as an input parameter,
here we have to measure the stress. In this study of the
stress-triaxiality we are interested in both mean and shear stresses.
Therefore the whole stress tensor $\sigma_{\alpha \beta}$ is
needed. The stress tensor (the negative of the pressure) can be
calculated atomistically on each time-step using the virial
formula:~\cite{allen}
\begin{equation}
\sigma_{\alpha \beta} = - \frac{1}{V} \left( \sum_i p_{i \alpha} 
p_{i \beta} / m_i  + \sum_i \sum_{j>i} r_{ij \alpha} f_{ij \beta} \right).
\label{eq_stresstensor}
\end{equation}
The first term in the stress tensor is the kinetic contribution of
atoms denoted with $i$ and having masses $m_i$ and momenta $p_i$. The
second term, a microscopic {\it virial} potential stress, consists of
sums of interatomic forces $f_{ij}$ of atom pairs $\langle ij\rangle$
with corresponding distances $r_{ij}$.  It should be noted, that here
and in the rest of the Article $i$ and $j$ denote the atoms, and
$\alpha$ and $\beta$ the Cartesian coordinates.  Note that the thermal
stress is included, although in practice in these simulations it
contributes less than 1 GPa, less than 10\% of the yield stress value,
and never dominates the changes in stress.

\subsection{Simulations Performed}
\label{simulations}

Typically in the simulations carried out here, the cube consists of 60
FCC unit cells in each direction, giving 864~000 atoms. The
equilibrium side-length of such a copper system is $l=21.6$~nm at room
temperature and ambient pressure.  The radius of the spherical void
cut from the system, unless otherwise noted, is 0.1 of the side-length
of the box; thus $2.2$~nm. After the void is cut, there are 860~396
atoms in the system.

The relatively inexpensive potential used enables us to do extensive
simulations in time.  A single time-step takes typically about 40 sec
of CPU-time in a system with 864~000 atoms in a Linux workstation with
Intel Xeon 1700MHz processor. The longest calculation required 835~050
time-steps corresponding to 5.6~nanoseconds. The time-step was $6.7$
femtoseconds.

As mentioned earlier, in order to study the effects of the
stress-triaxiality and different modes of expansion on the
void-growth, we have applied three different types of expansion,
namely uniaxial, biaxial, and triaxial. The strain-rates used for each
of the three modes of expansion are $\dot{\varepsilon} = 10^{10}$/sec,
$10^9$/sec, $5 \times 10^8$/sec, $10^8$/sec, and $10^7$/sec. For the
lowest strain-rate, the MD code was parallelized in order to take
advantage of massively parallel computers. The parallelization was
done using a spatial domain-decomposition, and was shown to scale
nearly linearly up to 128 processors.  The parallel code was used in
the case with 835~050 time-steps mentioned above, for example.

For comparison in the elastic regime, we have also performed
simulations without a void in all three modes of expansion.  These
simulations have been used to determine the bulk, elastic
stress-strain response of the EAM copper and hence the elastic
constants.  Without a void, the system is not so strain-rate and
system size dependent, at least up to the point of failure, so the
so-called ``no void'' simulations have been performed with a smaller
system size, 45 FCC-cells in each direction (364~500 atoms) and at the
single strain-rate $\dot{\varepsilon} = 10^9$/sec. A uniaxial study of
the 60$^3$ system size, but with a smaller initial void radius of
$1.1$~nm, was carried out with the strain-rate $\dot{\varepsilon} =
10^8$/sec in order to study the void-size dependence.  In this case
the system with the void contains 863~543 atoms.

It should be mentioned, too, that all of the intermediate strain-rate
simulations ($\dot{\varepsilon} = 10^8$/sec and $5 \times 10^8$/sec)
expansion were not started from equilibrium conditions at $P = 0$~MPa,
but from systems expanded previously at the strain-rate
$\dot{\varepsilon} = 10^9$/sec.  These simulations have been restarted
well before yielding, when the system's behavior is rate independent,
and relaxed for 2000 time-steps, or 13.4 picoseconds, without
expansion before continuing the expansion at the intermediate strain
rates. The energy is conserved during the relaxation in MD
simulations. These restarts have been accomplished at strain values
$\varepsilon = 4.12$~\%, $\varepsilon = 2.06$~\%, and $\varepsilon =
1.72$~\% in uniaxial, biaxial, and triaxial cases, respectively.

\section{Stress-Strain Behavior and Stress-Triaxiality}
\label{stressstrain}

Let us begin to explore the results of the MD simulations by looking
at the stress-strain curves. Figure~\ref{fig1} shows these curves for
each of the modes of expansion at all the strain-rates computed. The
data from ``no void'' cases are also plotted. The mean or hydrostatic
stress,
\begin{equation}
\sigma_m = \frac{1}{3} \, \mathrm{Tr} \, \mathcal\sigma_{\alpha \beta},
\label{eq_meanstress}
\end{equation}
is plotted to indicate the principal impetus for void growth.  Note
that the strain is the {\it engineering strain}, defined as the
expanded system size divided by the original system size minus one. In
the uniaxial and biaxial cases the strains are the principal strain
values $\varepsilon$ in the direction of the strain, such as
$\varepsilon_x = \varepsilon$, $\varepsilon_y = \varepsilon_z = 0$ in
the uniaxial case, and $\varepsilon_y = \varepsilon_z = \varepsilon$,
$\varepsilon_x = 0$ in the biaxial case. Hence the strain is the value 
of a non-zero diagonal term of $t \dot{\mathcal E}$ in Eq.~(\ref{eq_update}).  
The mean strains  $\varepsilon_m$ are 1/3 and 2/3 of the plotted uniaxial 
and biaxial strains, respectively.
Thus the total volumetric strain-rates are not the same in the
different expansion modes. In the triaxial cases the plotted and the
mean strain values are the same. The shape of the stress-strain curves
do not differ much depending on the modes of expansion in these cases,
at least when plotted versus mean strain. Independent of the
strain-rate, and whether with or without a void, the stress-strain
curves lie essentially on top of each other during elastic expansion,
{\it i.e.}, the initial smooth behavior when the system is still
recoverable and has not deformed plastically.

The stress-strain curve starts to deviate from the trend of the
elastic behavior at a specific, ``critical'' point which we call here
a {\it yield point}. In other quantities we measure a change in
behavior happens at a specific point, too, and as we shall see later
the critical or yield points mostly coincide with each other, {\it
i.e.}, their strain values are approximately the same independent of
from which quantity we derived it. The same point is also the one when
the void starts to grow, which is the primary mechanism for plasticity
in this study.  Here we define numerically the yield point of the
cases with void by comparing their stress-strain curves with the
reference ``no void'' curve, which behaves elastically beyond the
yield points of the other cases [cf.\ Fig.~\ref{fig1}(a) inset].
Ultimately the no void case does fail by homogeneous nucleation
of voids, and this is the reason for the drop in the mean stress. 
There is a small offset between cases with a void compared to the case
without a void due to the elastic relaxation of the void.  The value
of the stress at the yield point in the cases with void is lower with lower strain-rate, and
thus the strain to yield is also lower.  In each of the modes of
expansion, the stress at the yield point for the strain-rate
$\dot{\varepsilon} = 10^7$/sec is close to the value to which the
higher strain-rates converge. Of course, at much lower strain rates
the physics changes, new mechanisms become active, so this value need
not hold for arbitrarily low strain rates.  However, it is noteworthy
that the stress at the yield point is not scaling with strain, contrary to the
experimental finding for the spall strength explained in
Section~\ref{intro}. Overshooting, the phenomenon that the maximum
stress is much higher than the stress at the yield point, is evident
here for the higher strain rates.  The scaling of the spall strength
versus the strain-rate~\cite{Kanel} with an exponent 0.2 is reproduced
here when one compares the maximum stress values instead of the
stresses at the yield point for strain-rates $\dot{\varepsilon} =5
\times 10^8$/sec, $10^9$/sec, and $10^{10}$/sec, since then the exponent is $0.14-0.18$,
lowest for the uniaxial case and highest for the triaxial case. On the
other hand the stress value at the yield point, which is at the same
time the maximum stress, when $\dot{\varepsilon} = 10^7$/sec is very
close to the value of 6-8~GPa the spall strength scaling predicted
from the lower strain-rates mentioned in Section~\ref{intro}. It
should be noted also, that since we are limited to finite, fairly
small, system sizes, at late stages of the stress-strain curves, at
the failure, the data is not realistic anymore. The reason is that at
the plastic part of the stress-strain behavior when the void grows, it
also emits dislocations, and in a finite system with periodic
boundaries, when the dislocations have traveled long enough, they
propagate through the boundaries and reenter from the other side. In
the picture where we have a periodic array of voids in an infinite
system this means that the voids are so close to each other that they
start to interact. In reality voids are never arranged in a perfect
cubic lattice structure and in symmetric positions with respect to
each other, and thus the interactions of the voids in the simulations
with their periodic images are just an unphysical finite size effect.

In the shear stress or more precisely in the deviatoric part of the
stress tensor $\sigma_e$ plotted in Fig.~\ref{fig2}, a much bigger
difference is seen between the modes of expansion than in the mean
stress.  For the deviatoric part of the stress $\sigma_e$ we use {\it
von Mises stress}:
\begin{equation}
\sigma_e = [ 3 \, J_2 ]^{1/2},
\label{eq_misesstress}
\end{equation}
where $J_2=\frac{1}{2}{\mathrm{Tr}}~\sigma ^{\prime 2}$ is the second
invariant of the stress deviator $\sigma'_{\alpha \beta} =
\sigma_{\alpha \beta} - \sigma_m {\mathbb I}$.~\cite{Dieter} Thus von
Mises stress reads:
\begin{equation}
\begin{array}{ccl}
\sigma_e  &=& \left[ 3 \left(\sum_{\alpha > \beta} \sigma^2_{\alpha \beta}
- \sum_{\alpha > \beta} \sigma'_{\alpha \alpha} \sigma'_{\beta
  \beta} \right) \right]^{1/2} \\
&=& \left[ \frac{1}{2} \sum_{\alpha > \beta} (\sigma_{\alpha \alpha} 
-  \sigma_{\beta \beta})^2 + 3 \sum_{\alpha > \beta} 
\sigma_{\alpha \beta}^2 \right]^{1/2}.
\end{array}
\label{eq_misesstress2}
\end{equation}
While the mean stress at the yield point gets a value of about $\sigma
= 5.6-6.4$~GPa when loaded with strain-rate $\dot{\varepsilon} =
10^7$/sec in each of the three modes of expansion, von Mises stress
has a value of $\sigma_e =2.0$~GPa and 0.7~GPa in the uniaxial and
biaxial cases, respectively. In the triaxial case it should be zero by
symmetry, and the difference from zero, representing symmetry-breaking
effects, is small.  Thus the loading differences between the modes of
expansion are quantified in von Mises stress. After the onset of
plasticity or the void growth, von Mises stress gets a value of about
$\sigma_{e} =0.4$~GPa, 0.2~GPa, and 0.1~GPa, in the uniaxial, biaxial,
and triaxial cases, respectively, independent of the strain-rate, but
with significant fluctuations in this regime.  In this period also
dislocations move under the action of the shear stress until the
stress has dropped to the point that it is no longer sufficient to
move a dislocation through the forest of dislocations. The final value
of the shear stress corresponds to the flow stress, and they are close
to the tensile strength values of copper, 200-400~MPa, quoted in the
literature.~\cite{budinski}

Although von Mises stress captures differences between the loading
modes quite well, an even better quantity to study is the ratio
between hydrostatic and shear stresses, the {\it stress-triaxiality}
\begin{equation}
\chi = \sigma_m / \sigma_e,
\label{eq_triax}
\end{equation}
which has been plotted in Fig.~\ref{fig3}.  In the uniaxial case the
stress-triaxiality starts from the value $\chi \simeq 3.0$ and slowly
decreases linearly to a value $\chi \simeq 2.8$ until the onset of
rapid growth at the yield point. After the rapid increase the
stress-triaxiality saturates at $\chi \simeq 11.0-16.0$.  The
stress-triaxiality in the biaxial case starts with a much larger value
than in the uniaxial case. It begins at $\chi \simeq 6.0$ and
increases linearly to a value $\chi \simeq 8.0$ at the yield point
where it grows rapidly to a value of $\chi \simeq 15.0-30.0$. 
We have noted the correspondence of the final von Mises stress and 
the flow stress above. Similarly the spall strength provides an
experimental measure of the mean stress that can be supported in 
void growth. Thus the stress-triaxiality
values can be compared with the ratio of the spall and the tensile
strength of copper. Previously quoted literature values for them are
6-8~GPa and 200-400~MPa, respectively, giving for their ratio values
between 15 and 40, and thus comparable with the stress-triaxiality
values here.  The comparison is not fully rigorous, but it
provides an indication of how reasonable the final stress triaxiality
values are in terms of experiment. 
Since the stress-triaxiality is the mean stress divided
by von Mises stress, which is equal to zero in the triaxial case until the
yield point and very small even after that, the stress-triaxiality is
diverging and therefore not plotted here in that case.  The
stress-triaxiality values in the uniaxial and biaxial cases at the
elastic part of the simulation is compared here also with the values
one get from the elasticity theory:~\cite{eshelby}
\begin{equation}
\chi = \frac{1}{3} \, \frac{C_{11} + 2 C_{12}}{ C_{11} - C_{12}} \, \, \Xi. 
\label{anal}
\end{equation}
$\Xi = 1$ in the uniaxial case and $\Xi = 2$ in the biaxial case.  The
literature values for the elastic constants of copper are $C_{11} =$
168~GPa and $C_{12} =$ 121~GPa.~\cite{AM} Thus $\chi =2.9$ and $\chi
=5.8$ in the uniaxial and biaxial cases, respectively, which compare
quite well with the simulations presented here.  When the elastic
constants are derived from the stress-strain curves as $\varepsilon
\to 0$, they are close to the actual experimental values:
$C_{11}\simeq$ 162~GPa, $C_{12} \simeq$ 121~GPa and $C_{11}\simeq$
168~GPa, $C_{12} \simeq$ 124~GPa in the uniaxial and biaxial cases,
respectively.

\begin{table*}
\caption{\label{tab1} The onset of plasticity associated with void
growth, as indicated by 3 different criteria: deviation from elastic
behavior in the mean stress, von Mises stress and stress triaxiality.
Their threshold values, together with the corresponding strain values,
are tabulated for uniaxial, biaxial and triaxial expansion at the
strain rates $\dot{\varepsilon}=10^9$/sec, $5\times10^8$/sec, $10^8$,
and $10^7$/sec.  In particular, the third and fourth columns show the
mean stress values $\sigma$ and the corresponding mean engineering
strain values $\varepsilon_m$, respectively, at the critical or yield
point at which the mean stress first deviates from the elastic
stress-strain curve.  Analogously, the fifth and sixth columns show
the yield point as indicated by von Mises stress $\sigma_e$ and the
corresponding mean engineering strain values $\varepsilon_m$,
respectively.  The seventh and eighth columns show the yield point as
indicated by the stress triaxiality $\chi$ and the corresponding mean
engineering strain values $\varepsilon_m$, respectively.  Note the
small but significant differences in the yield point as indicated by
these three different criteria.  The error bars of the values are of
the order of last reported digit. The details of the simulations are
in the caption of Fig.~\ref{fig1} and the curves, from which the yield
data have been calculated, are plotted in Figs.~\ref{fig1}-\ref{fig3}.
Note that, as expected, von Mises stress is small and erratic in the
triaxial case, so those von Mises and stress-triaxiality data are not
tabulated.}
\begin{ruledtabular}
\begin{tabular}{l c|c c|c c|c c}
$(\varepsilon_x, \varepsilon_y, \varepsilon_z)$& $\dot{\varepsilon}$ (sec$^{-1}$) & $ \sigma$ ( GPa ) & $\varepsilon_m$ (\%) 
& $\sigma_e$ ( GPa ) & $\varepsilon_m$ (\%)& $\chi $ & $\varepsilon_m$ (\%) \\
\hline
$(\varepsilon,0,0)$ & $10^9$ & 5.87 & 1.85 & 2.12 & 1.87 & 2.78 & 1.92 \\
$(\varepsilon,0,0)$ & $5\times10^8$ & 5.82 & 1.84 & 2.09 & 1.84 & 2.79 & 1.84 \\
$(\varepsilon,0,0)$ & $10^8$ & 5.65 & 1.77 & 2.01 & 1.77 & 2.80 & 1.77 \\
$(\varepsilon,0,0)$ & $10^7$ & 5.60 & 1.77 & 2.00 & 1.77 & 2.79 & 1.77 \\
 & & & & & & \\
$(0,\varepsilon,\varepsilon)$ & $10^9$ & 6.50 & 2.02 & 0.79 & 2.02 & 8.29 & 2.02\\
$(0,\varepsilon,\varepsilon)$ & $5\times10^8$ & 6.50 & 2.02 & 0.79 & 2.02 & 8.23 & 2.02 \\
$(0,\varepsilon,\varepsilon)$ & $10^8$ & 6.03 & 1.85 & 0.75 & 1.87 & 8.08 & 1.86\\
$(0,\varepsilon,\varepsilon)$ & $10^7$ & 5.96 & 1.83 & 0.74 & 1.82 & 8.02 & 1.82 \\

 & & & & & & \\
$(\varepsilon,\varepsilon,\varepsilon)$ & $10^9$ & 7.25 & 2.30 & & & & \\
$(\varepsilon,\varepsilon,\varepsilon)$ & $5\times10^8$ & 7.25 & 2.30 & & & &  \\
$(\varepsilon,\varepsilon,\varepsilon)$ & $10^8$ & 6.50 & 2.00 & & & & \\
$(\varepsilon,\varepsilon,\varepsilon)$ & $10^7$ & 6.33 & 1.94 & & & & 
\end{tabular}
\end{ruledtabular}
\end{table*}
The critical mean stress, von Mises stress, and the stress-triaxiality
values where their behaviors start to deviate compared to the elastic
ones, or what we call yield points, are summarized in Table~\ref{tab1}
for the strain-rates $\dot{\varepsilon}=10^9$/sec, $5\times10^8$/sec,
$10^8$, and $10^7$/sec of the principal strains. In the case of the
highest strain-rate $\dot{\varepsilon}=10^{10}$/sec, the shapes of the
stress-strain curves are so much rounded due to over-shooting, that
there is no clear point, where the stress-strain curve deviates from
the elastic behavior, and thus our definition of the yield point is no
longer suitable.  In comparing the mean strain values $\varepsilon_m$
(as in the Table) at the onset of plasticity for a particular strain
rate, one finds that the uniaxial expansion always starts to yield at
the least strain, and the hydrostatic expansion, at the greatest
strain. There are two effects that contribute to the increase in the
plastic threshold as the triaxiality increases.  First, the shear
component of the applied stress contributes to the resolved shear
stress and lowers the threshold for heterogeneous nucleation of
dislocations at the void surface.~\cite{unpubl} And second, the
volumetric strain-rate is lowest in the uniaxial case and the highest
in the hydrostatic case. Strain-rate hardening then leads to an
increase in the stress value at the onset of plasticity as the
triaxiality increases.  The difference between the critical strain
values when defined as when a behavior deviates from the elastic
behavior is nearly negligible and thus independent of whether one uses
the criterion from the hydrostatic stress, von Mises stress or
stress-triaxiality curves.  The differences reflect mainly the
difficulties in defining the point what we call the yield
point. However, we will see later that if the mean stress and von Mises
stress start to deviate from the elastic behavior with the same ratio
as they have during elastic expansion, the stress-triaxiality may
deviate a bit later than the other quantities.  We shall see later,
too, that the onset of plasticity defined from these quantities is
very close to where the void starts to grow.

\section{Plastic strain and plastic work}
\label{plasticstrain}

After sufficient expansion, the system yields and the mean stress is
observed to drop with respect to the elastic response.  Then as the
simulation box continues to expand, the stress remains roughly
constant until the precipitous drop at final failure.  In the region
of increasing strain but roughly constant mean stress, most of the
strain is in the form of plastic strain, a macroscopic measure of the
plastic, permanent and irrecoverable deformations in the system.  In
this Section we study the macroscopic quantities of plasticity such as
mean and equivalent plastic strains as well as the plastic work, and
in the next Section in more detail what are the actual plastic
deformations visible in the void. The concomitant dislocations related
to void's shape and volume changes are studied
elsewhere.~\cite{belak3,rudd,unpubl}

In deriving the plastic strain here it is assumed that the tetragonal
symmetry is approximately preserved and thus the off-diagonal terms of
the stress tensor are negligible.  Following the literature we
separate the strain increment $d\varepsilon = \dot{\varepsilon} \, dt$
into elastic and plastic parts.~\cite{hill} Thus by definition the
plastic strain increment becomes
\begin{equation}
\dot{\varepsilon}^P_{\alpha \beta} \, dt = 
\dot{\varepsilon}^{tot}_{\alpha \beta}  \, dt 
- \dot{\varepsilon}^E_{\alpha \beta} \, dt,
\label{eq_plastic}
\end{equation}
where $\dot{\varepsilon}^{tot}_{\alpha \beta} \, dt$ is the total
increment of the strain. Below we use $\dot{\varepsilon}$ instead of
$\dot{\varepsilon} \, dt$ since $dt$ can be divided from both sides of
Eq.~(\ref{eq_plastic}). The total strain increment is an input
parameter in these strain-controlled simulations.  It is given by the
strain-rate matrix $\dot{\mathcal E}$. The compliance ${\mathcal S}$
relating the elastic strain increment to the stress increment is
derived from the stress-strain curves in the elastic region by:
\begin{equation}
\dot{\varepsilon}^E_{\alpha \beta}(\sigma_{\alpha \beta}) = {\mathcal S} 
\dot{\sigma}_{\alpha \beta}.
\label{eq_eepsilon}
\end{equation}
The stress matrix $\sigma_{\alpha \beta}$ is calculated each time-step
using Eq.~(\ref{eq_stresstensor}). The elastic compliance tensor
$S(\sigma_{\alpha \beta})$ is retrieved from the elastic part of the
stress-strain curve of the cases without the void as follows. Due to
the nonlinearity of a stress-strain curve we have not only retrieved
the slope of it, which would give $3B = C_{11} +2C_{12}$, where $B$ is
the bulk modulus, but fitted a fourth order polynomial to the strain
versus stress curve, whose derivative gives us $1/3B^{-1}(\sigma_m)$.
This is done separately for the uniaxial, biaxial, and triaxial
no-void cases, and the respective curves are used for the cases with
the void.  It should be mentioned, too, that in the derivation of the
bulk modulus the mean total logarithmic strain is used instead of the
engineering principal strain used in the plots of this Article.
Similarly the term $C' = \frac{1}{2} \left(C_{11} - C_{12}\right)$ is
derived using a fourth order polynomial in the mean strain versus von
Mises stress curve giving $1/C'$.  Note, that when $C'$ is derived
from the plot using mean strain there are prefactors 1/3 and 2/3 for
$1/(C_{11} - C_{12})$ in the uniaxial and biaxial cases, respectively.
Using formulas which relate $S_{11}$ and $S_{12}$ to $C_{11}$ and
$C_{12}$ in cubic crystals,~\cite{Dieter} and the correspondence
between elastic constants and moduli we get for $S_{11}$ and $S_{12}$
\begin{equation}
S_{11}  = \frac{1}{9B}+\frac{1}{3C'}, \,\,\, 
S_{12}  = \frac{1}{9B} -\frac{1}{6C'}.
\label{eq_ss}
\end{equation}
Using the $S_{11}(\sigma_m)$ as $S_{\alpha \alpha}(\sigma_m)$ and
$S_{12}(\sigma_m)$ as $S_{\alpha \beta}(\sigma_m)$ due to the symmetry
and neglecting off-diagonal terms, which are small compared to the
diagonal ones, we get all the necessary terms for $S(\sigma_m)$, and
thus $\dot{\varepsilon}^E_{\alpha \beta}(\sigma_m)$ from
Eq.~(\ref{eq_eepsilon}). Note, that since the $\sigma_m$ is used as a
parameter instead of $\sigma_{\alpha \beta}$, the von Mises stress must be
mapped with the mean stress when finding the corresponding $C'$. This
was done again by fitting the von Mises vs.\ mean stress curves with
fourth order polynomials.

Subtracting the elastic strain from the total strain as in
Eq.~(\ref{eq_plastic}) we get the mean plastic strain increment
\begin{equation}
\dot{\varepsilon}^P_m = \frac{1}{3} \sum_\alpha \dot{\varepsilon}^P_{\alpha \alpha},
\label{eq_mean_plastic}
\end{equation}
which time-integral is plotted in Fig.~\ref{fig4} for all the loading
modes and strain-rates. In these plots one sees that after the yield
point the mean plastic strain first increases roughly exponentially,
although the region is too small to be definitive, and thenceforth
roughly linearly. Note that the mean plastic strain is not the
equivalent plastic strain commonly used in plasticity, which will be
defined below, but a measure of the porosity. This will be studied in
the next Section, where the mean plastic strain will be compared with
the growth of the volume of the void.

We turn now to the quantification of the dislocation flow,
conventionally computed at the continuum level as the second
invariant of the deviatoric plastic strain, the {\it equivalent 
plastic strain}.  Typically in the case of tetragonal total strain,
the equivalent plastic strain rate would be calculated as:
\begin{equation}
\dot{\varepsilon}^P_e  = \frac{1}{3} \left\{ 2
\sum_{ \alpha > \beta}  \left(\dot{\varepsilon}^P_{\alpha \alpha}
- \dot{\varepsilon}^P_{\beta \beta} \right)^2  \right\}^{1/2}.
\label{eq_equivalent}
\end{equation}
The equivalent plastic strain is calculated in turn as
\begin{equation}
\varepsilon^P_e (t) = \int_0^{t} \dot{\varepsilon}^P_e(t') \, dt'.
\label{eq_equivalent_t}
\end{equation}
In practice this formula for the equivalent plastic strain is
problematic in MD for several reasons.  First, the time and length
scales in MD are much shorter than those assumed in continuum
formulations of plasticity.  The time scale is a problem because
dislocation flow becomes partially reversible at short enough time
scales.  Thermal fluctuations cause reversible oscillations of
dislocations and fluctuations in the local elastic strain.  To the
contrary, the integrand in Eq.\ (\ref{eq_equivalent_t}) is positive
definite, as appropriate for plastic deformation that is cumulative
even when reversed.  In practice, the application of Eq.\
(\ref{eq_equivalent_t}) in MD gives a result dominated by the
fluctuations for small time increments; in fact, in our attempt to
apply the formula to the MD deformation every 10 time steps, the
contribution of the fluctuations was 22 times as large as the applied
total mean strain (these values are obtained from the biaxial case
with strain-rate $\dot{\varepsilon} = 10^8$/sec). The formula must be
modified to be insensitive to thermal fluctuations.  Second, the
formula for the equivalent plastic strain assumes isotropic plasticity
in the following sense.  In isotropic plasticity, the plastic flow is
driven by the shear stress quantified by the von Mises stress.  The
equivalent plastic strain is conjugate to the von Mises stress, and
therefore takes on a particular signficance in the theory.  Implicit
is the assumption, for example, that slip systems that experience the
same shear stress will exhibit the same plastic strain.  This
assumption is violated in MD for two reasons.  Once again, the thermal
fluctuations may cause the initiation of flow on one glide system
before that on a symmetrically related system.  This effect is
observed in our MD simulations.  Typically the symmetry is restored
after a brief period, but because the plastic strain is cumulative,
the symmetry breaking fluctuation is never eliminated from the plastic
strain (\ref{eq_equivalent_t}).  Second, a more mundane reason the
assumption of isotropy fails is that the single crystal systems are
anisotropic, both because of the specific glide planes involved and
because of the elastic constants, especially in copper.

It may be possible to rectify these problems while retaining the
basic formulation of the equivalent plastic strain, for example 
through a suitable multi-resolution calculation of the integral 
(\ref{eq_equivalent_t}).  We have made several attempts at a
new formulation, but we were not able to develop a satisfactory
algorithm, providing a meaningful measure of the plastic strain
on MD time scales based on the equivalent plastic strain integral 
(\ref{eq_equivalent_t}).  We found that we could eliminate the
anomalies due to fluctuations or the anisotropy, but not both
simultaneously in a robust manner.

We have therefore turned to a different quantification of the plastic
strain.  Certainly, the full deviatoric plastic strain tensor is a
measure of the plastic flow, conjugate to the deviator stress.  Its
rate of increase is given by the traceless part of
Eq.~(\ref{eq_plastic}).  Typically, the rate would be integrated in a
cumulative fashion, but we will not do so.  The nature of our
simulations is such that at the continuum level plastic flow is only
expected in one direction, so any sign reversal may be attributed to
fluctuations.  We then calculate
\begin{equation}
\varepsilon^P_{\alpha \beta} (t) = \int_0^{t} 
\dot{\varepsilon}^P_{\alpha \beta}(t') \, dt',
\label{eq_equivalent_t2}
\end{equation}
where the plastic strain rate is given by Eq.\ (\ref{eq_plastic}).
We emphasize again that no absolute value is taken, so
fluctuations cancel.

Then in order to have a scalar quantification of the plastic strain,
we compute the $J_2$ invariant, normalized as the equivalent plastic
strain would be:
\begin{equation}
\varepsilon^P_e (t) = \frac{1}{3} \left\{ 2 
\sum_{ \alpha > \beta}  \left(\int_0^t \dot{\varepsilon}^P_{\alpha \alpha} 
(t') \, dt' - \int_0^t \dot{\varepsilon}^P_{\beta \beta} (t') \, dt' \right)^2
 \right\}^{1/2}.
\label{eq_equivalent2}
\end{equation}
We must stress that this quantity is not equal to the equivalent
plastic strain commonly used in plasticity, except in the
extraordinary case of monotonic isotropic plasticity.  It is not
conjugate to the von Mises stress, for example, in our MD simulations.
Nevertheless, it is a useful qualitative measure of the degree of
plasticity, and it allows us to compare the plastic response as the
system is loaded in different ways and we call it here {\it equivalent
plastic strain} for simplicity.

The evolution of the equivalent plastic strain during uniaxial and
biaxial expansion is plotted in Fig.~\ref{fig5}.  In the triaxial
case it is essentially zero, as expected by symmetry, and therefore it
has not been plotted.  In practice, the stress during triaxial
expansion has only a negligibly small fluctuating shear component, so
the calculated elastic shear strains are very small, too. Since
neither the box strain nor the elastic strain has an appreciable shear
component, the equivalent plastic strain is found to be zero.

Now, once the tensors for both the stress and the plastic strain are
derived, (actually only the diagonal terms of the plastic strain are
needed), the plastic work can be calculated,
\begin{equation}
W_P(t) = \sum_\alpha \int \dot{\varepsilon}^P_{\alpha \alpha} 
\sigma_{\alpha \alpha} \, dt,
\label{eq_plastic_work}
\end{equation}
(see Fig.~\ref{fig6}).  It should be compared with the temperature
from the same simulations, Fig.~\ref{fig7}.  Note, that in these
simulations, when the dilational strain is applied, the thermostat is
turned off and thus the temperature is allowed to change.  First the
system cools in the elastic regime due to adiabatic cooling on
expansion, but when plastic deformations begin, work is done in the
system resulting in heating. We find that the increase in plastic work
does not match exactly with the temperature.  In principle we expect
several effects to contribute to this difference: the surface energy
of the void, the defect formation energies for dislocations and point
defects, further adiabatic cooling, and any error in calculating the
elastic energy or strain from the stress.  Using the best data available to
bound the contributions from surface, defect and adiabatic cooling
energies, we find that there remains an energy deficit that we
attribute to an error in the calculated elastic energy.  The error
comes from the use of the average stress despite stress inhomogeneity
in the system due to the void: in the plastic work the product of
plastic strain and stress is calculated with averaged quantities,
while in the temperature the product is calculated at level of each
atom and averaged afterward.

\section{Void Evolution}
\label{voidevol}

\subsection{Growth of the Void}
\label{voidvol}

We now consider the volume and shape evolution of the void.  During
the MD simulations undergoing expansion, the surface of the void is
determined by finding individual atoms that belong to the
surface. This is done by creating a fine two-dimensional mesh, in
which each mesh point corresponds to spherical angular coordinates
$(\phi, \theta)$. An atom is found to represent the surface at each
point of the mesh, with some atoms representing multiple mesh points.
In particular, taking the origin to be the center of the void, within
the solid angle associated with each mesh point, the atom that is
closest to the origin is defined to be the surface atom at that mesh
point. There are, however, some uncertainties related to this
method. If the mesh is too dense with its size diverging it can
capture almost all the atoms in the system. On the other hand if it is
too sparse, it may neglect some surface atoms, especially when void is
anisotropic, non-spherical, and has some sharp edges in it. Therefore
we introduced a width to each of the atoms by drawing a circle around
it that implies a width $(d\phi, d\theta)$ to the angles, so that one
atom can occupy several mesh points in a fine mesh. We have typically
75-100 points for each angular coordinate, giving a total of
5625-10000 mesh points. In the surface of the void there are typically
few thousand atoms. Besides introducing the width to the atoms, we
also select atoms based on their radial distance: if an atom has much
larger radial distance $r$ compared to its neighbors it is neglected
in order not to capture atoms that do not belong to the surface.

Once the surface atoms are identified, the surface is tessellated
using a generalization of the Delaunay triangulation
method.~\cite{Sloan} The Delaunay triangulation is an optimal
triangulation of a collection of points--in our case atoms--on the
plane.  It is optimal roughly in the sense that the aspect ratio of
the triangles is as near to unity as possible; more precisely, the
Delaunay theorem guarantees that there is a unique (up to degeneracy)
triangulation such that if any triangle in the triangulation is
circumscribed by a circle, none of the other points will be in the
interior of the circle.  The Delaunay theorem, as formulated, does not
apply to points on a curved surface.  In fact, there appears to be a
topological obstruction to the existence of a unique, optimal
triangulation on a closed surface when the Euler character is
non-zero.  Nevertheless, it is possible to extend the Delaunay
triangulation algorithm to achieve a locally optimal triangulation
almost everywhere.  The approach we have taken is to project the
points patchwise onto flat surfaces.  In particular, stereographic
projections are used to project the upper and lower hemispheres
separately onto planes.  Cylindrical coordinates are used to project
the equatorial region to a cylinder.  The Delaunay algorithm is used
to triangulate each of these projected regions.  The patches overlap
at latitudes of $\pm45^{\circ}$ where the projections are not too
distorted.  The three patches are sewn together using a simple
advancing front triangulation at the boundaries.

Using this triangulation, the volume of the void can be calculated
precisely by summing up the volumes of the tetrahedra with one apex at
the center of the void and the opposing face on the void surface.  As
we shall see below, the void shape evolves to be far from
spherical. Therefore the approximation of the void surface by
triangles captures the shape better than just assuming it to be
spherical and using only the solid angles and radial distances of a
sphere, when calculating the volume of the void. An advantage of this
method is that if some atom, which should be taken into account, is
missing from the surface of the void, its position is filled with the
triangles created by its neighboring atoms, and thus the ``hole'' is
well approximated by its neighbors.

In Fig.~\ref{fig8} the porosity or the ratio between the volume of
the void and the total volume of the system,
\begin{equation}
f = V_{void}/V,
\label{eq_void_frac}
\end{equation}
is plotted for a fraction of the simulations of the strain-rates
$\dot{\varepsilon} = 10^{10}$/sec, $10^9$/sec, $5 \times 10^8$/sec,
and $10^8$/sec.  It should be noted that in order to get information
about the positions of the atoms for the strain values of the interest
({\it i.e.}, close to and after the onset of plasticity) the
calculations were restarted from already expanded system. After the
restart the expansion was applied with the same strain-rate as earlier
but now to the already expanded system, thus the strain-rate was
increased by a few percent compared to the original [since ${\mathcal
H}_0$ in Eq. (\ref{eq_update}) was the restart value].  Therefore
these simulations are not precisely comparable with the ones plotted
in Figs.~\ref{fig1}-\ref{fig7}, where the continuum quantities are
shown. In these plots, as for the mean plastic strain, in most cases
first the void grows exponentially and then (if the calculation has
been carried out that long) it changes to a linear growth, see
especially Fig.~\ref{fig8}(b) and the strain-rates $10^9$/sec and $5
\times 10^8$/sec there.  The cross-over to the linear growth happens
at the same point as the rate of decrease of the mean stress slows.
Although it is beyond the scope of this Article to go into the
analysis, the reduction in the growth rate coincides with the point at
which the dislocation density along the shortest distance between the
void and its periodic image (at the apices of the faceted void, cf.\
Section~\ref{voidshape}) reaches saturation, and the nature of the
dislocation activity changes dramatically.  This can be interpreted as
a finite size effect as the void approaches the boundary of the
simulation box or as a start of the coalescence process of the void 
with its own periodic image. The void-void interactions and the 
coalescence  process of two voids in a less restrictive geometry will be 
presented elsewhere.~\cite{unpubl2}
The shapes of the porosity curves as a whole can be
compared with the mean plastic strain plotted in
Fig.~\ref{fig4}. Although the volume of the void is not calculated
throughout the whole simulation, one sees easily, that there is
correspondence between mean plastic strain and the volume of the
void.  There is of course an offset at the elastic part
of the simulations, since mean plastic strain equals zero then, but
the initial volume of the void is finite. The correspondence will be
revisited and studied more carefully in Section~\ref{summary}.
However, it can be concluded already here that the macroscopic
quantity mean plastic strain captures the microscopic behavior of the
void growth very well. Effects such as the excess volume associated
with defects are negligible.  This also means that the matrix material
is plastically incompressible, the dilation comes from the void
growth, and thus it is consistent with the Gurson type of continuum
models.~\cite{Gurson}

\subsection{Shape Evolution of the Void}
\label{voidshape}

Let us now look at the shape evolution of the void in more detail.  In
Fig.~\ref{fig9} snapshots of the void are shown from uniaxial
expansion at the strain-rate $\dot{\varepsilon} = 10^8$/sec.  There
are several interesting aspects in the snapshots. In the first two
snapshots at strains $\varepsilon = 5.05\%$ and $5.26\%$ when the
system still behaves elastically, the void is expanded in the
$x$-direction, which is the direction of the strain. However, after
that the void makes a rapid shape change and becomes more extended in
the transverse $y$ and $z$-directions, {\it i.e.}, the strain-free
directions, than $x$-direction. This prolate-to-oblate transition may
be counterintuitive, but the behavior has been observed previously in
continuum calculations,\cite{castaneda,budiansky,Andersson} and it has
been related to the appearance of shallow dimples in the fractography
studies of ductile fracture surfaces in low triaxiality conditions.
See also studies of non-spherical voids.~\cite{gologanu1,gologanu2}
For example, Budiansky {\it et al.}, Ref.~\onlinecite{budiansky},
investigated void shape change in
a non-linear viscous plastic flow model.  They explained the oblate
growth of voids under uniaxial loading as due to a non-linear
amplification of the shear stress on the surface of the void, with the
maximal void growth rate at the locations of maximal von Mises stress: the
equator.  Their analysis does not apply directly to our simulations
since they neglect elasticity, and the non-linear viscous solid model
they have used is not expected to be a precise description of the plastic
flow early in our MD simulation when the prolate-to-oblate transition
takes place.  Furthermore, it is not clear from our simulations what
value should be assigned to the strain-rate exponent that controls
the non-linearity in the model of Budiansky {\it et al}., although a
large value is reasonable.  Despite these differences,
the same localization of plastic flow to the equator of
the void and the transition to an oblate shape does appear in
both our simulations and the viscous solid model of Budiansky {\it et al}. 
Following some additional
expansion beyond the transition, the void begins to become faceted, as
visible in the last three snapshots. It should be mentioned that the
anisotropy visible in this uniaxial case is less pronounced in the
biaxial case. The cases with the hydrostatic loading are the most
isotropic and the octahedral shape, somewhat visible in
Fig.~\ref{fig9}(f), becomes more pronounced.~\cite{rudd} The
octahedral shape has been seen in spallation experiments in the FCC
metal aluminum,\cite{Stevens} and also in experiments on the
equilibrium void shape of another material, silicon, too, where it has
been used to calculate anisotropic surface energies through an inverse
Wulff construction.~\cite{Eaglesham} In void growth associated with
dynamic fracture in copper, several effects contribute to the
faceting: the low surface energy and high ad-atom energy of the
\{111\} surfaces common to FCC metals, the high anisotropy of the
copper elastic constants (A=3.21), and the \{111\} dislocation glide
systems.  These effects are analyzed in detail
elsewhere.~\cite{unpubl}

In order to characterize the shape change of the void not only
qualitatively and visually, but also quantitatively, multipole moments
of the void shape are calculated.  To the best of our knowledge, this
is the first time that multipole moments have been used to
characterize surface shape.  They are a powerful way to quantify the
evolution of the complex surface containing thousands of atoms, and
they are suitable for use in continuum models and experimental void
characterization as well.  Using spherical harmonics,
\begin{equation}
Y_{lm}(\vec{r}) \equiv Y_{lm} \{\theta(\vec{r}), \phi(\vec{r})\},
\label{eq_ylm}
\end{equation}
expressed as polynomials of Cartesian coordinates, in contrast to more
commonly used trigonometric forms,~\cite{jackson} we are able to
define different multipole moments of the void based on its surface
atoms:
\begin{equation} \begin{array}{lll}
Q_{lm} & \equiv &  \frac{1}{\bar{r}^2} \,
        \int \!  Y_{lm}(\theta,\phi) \, r^2(\theta,\phi) \, d\Omega,
\end{array}
\label{eq_qlm}
\end{equation}
where the mean square radius $\bar{r}^2 = \frac{1}{4 \pi} \int
r^2(\theta,\phi) \, d\Omega$.  This is in contrast to the volume
integral more commonly used when calculating multipole moments.  The
axial index of the moment ranges $m = -l,\ldots,l$, and for each $m$
except $m=0$, $Q_{lm}$ has both real and imaginary parts. Here we
concentrate on $l=1,2,3$, and 4. Only the positive moments of $m$ are
calculated, since the negative ones are related by
\begin{equation}
Q_{l,-m} = (-1)^m Q_{lm}^*. 
\label{eq_rel}
\end{equation}
In all 24 different terms are calculated.  The polynomial forms used
here of the moments are listed in the Appendix.

The moments $Q_{lm}$ are not rotationally invariant, but depend on the
way the coordinates $x$, $y$, and $z$ are chosen.  The set of $(2l+1)$
moments at fixed $l$ form an irreducible representation of the SO(3)
rotation group, and are mixed by rotations according to the usual
transformation rules.  They may be combined into a single rotationally
invariant combination for each $l$ according to
\begin{equation}
Q_l = \left[\frac{1}{2l+1} \sum_{m =-l}^{l} | Q_{lm} |^2 \right]^{1/2},
\label{eq_ql}
\end{equation}
see {\it e.g.} Ref.~\onlinecite{stein}. 
Their use drastically reduces the amount of data to be shown. Only the
positive $m$'s are needed for $Q_l$ due to the square of the norm of
$Q_{lm}$ and Eq.\ (\ref{eq_rel}).

Technically the calculation of the multipole moments has been done
using the information about the shape of the void obtained from the
surface triangulation procedure explained earlier.  As in the
calculation of the void volume, some refinements have been introduced
to reduce the uncertainty in the values of the moments that arises
from single atoms near the threshold for inclusion as surface atoms.
These borderline atoms can appear intermittently on the void surface
during the growth, and the tessellation is used to minimize their
effect on the moments.  In calculating the volume of the void the
triangulation gave one face of the tetrahedra that acted as small
volume elements for the total volume. Here the triangulation is used
to weight the atoms by the amount of solid angle associated with each
surface atom.  In particular, each triangle in the tessalation
contributes one third of its projected solid angle to each of the
three atoms that make up its vertices, where the solid angle of a
triangle is computed using the formula $\delta \Omega = A_1 + A_2 +
A_3 - \pi$, where $A_i = \arccos \left( \frac{c_{i}-c_{i+1}c_{i+2}}
{\sqrt{1-c_{i+1}^2} \sqrt{1-c_{i+2}^2}} \right)$ and $c_{i} =
\hat{x}_{i+1} \cdot \hat{x}_{i+2}$ for $i=1,2,3$ (mod 3) and where
$\hat{x}_{i}$ is the unit vector in the direction of the $i^{th}$
vertex of the triangle.~\cite{crc} The weight of each atom is the sum
of these solid angle contributions.  This reduces the sensitivity of
moments to the atomic discretization of the surface, since evanescent
atoms that occasionally appear and disappear from the fluctuating
surface only make a small, local change to the value of $r^2$.  It may
be of interest to note that in the course of the development of these
surface multipole moments, several different variations on the
definition of the moments were tried.  The definition presented here
(\ref{eq_qlm}) produced substantially less noise (up to a factor of 5
less noise) than the other definitions we tried, even though they all
showed the same trends in the shape evolution.  Using these weights
for the atoms and after normalizing each atom's $(x,y,z)$ coordinate
by its distance $r = (x^2+y^2+z^2)^{1/2}$ from the center of the void
all the terms in Eqs.~(\ref{eq_ql1})--(\ref{eq_ql4}), and
Eq.~(\ref{eq_ql}), are calculated. The center of the void is defined
to be the point where the three components of $Q_{1m}$, as given by
Eq.~(\ref{eq_ql1}), are zero.

Due to space limitations, only a fraction of the multipole moment data
is shown here. In Fig.~\ref{fig10}(a) the quadrupole moments $Q_{2m}$
for all the positive $m$ of the void are shown in the uniaxial case
for the strain-rate $\dot{\varepsilon} = 10^8$/sec.  This is the same
simulation as the snapshots in Fig.~\ref{fig9}.  Indeed, the
quadrupole moments are able to represent numerically the shape changes
one sees in the snapshots. In Figs.~\ref{fig9}(a) and (b) at strains
$\varepsilon = 5.05\%$ and $5.26\%$ the void is elongated to the
direction of the load, which is visible as $Q_{20} >0.$ Between
strains $\varepsilon = 5.47\%$ [Fig.~\ref{fig9}(c)] and $5.68\%$
[Fig.~\ref{fig9}(d)] the void is extended more transverse to the
direction of the strain, thus $Q_{20}<0$, and later it starts to
becomes more of octahedral shape and the absolute value of $Q_2$
saturates.

Figures~\ref{fig10}(b)--(d) show the rotational invariant multipole
moments $Q_l$, Eq.~(\ref{eq_ql}), in cases with uniaxial, biaxial, and
triaxial loading, respectively. Each of the cases has strain-rate
$\dot{\varepsilon} = 10^8$/sec.  In the plots it is clear that the
behavior that the quadrupole moment has first a non-zero value and
then makes a rapid dip but returns back to a non-zero value due to the
transverse elongation is the strongest in the uniaxial case. On the
other hand the octahedral shape measured by the $Q_4$ is more
pronounced in the biaxial and triaxial cases than in the uniaxial case
as explained qualitatively earlier. Hence we find that the multipole
moments introduce a good method to measure the shape changes of the
void. The non-zero values for $Q_3$, as well as $Q_2$ in other cases
than uniaxial, indicate that the void is not (cubically) symmetric in
these simulations. It should be mentioned that these first four
moments are enough to characterize the shapes of the void and the
higher moments contain little relevant information. This was checked
by creating a three dimensional surface based on the moment values and
drawing it in the same figure with the actual positions of the surface
atoms using a standard commercial visualization program. The surfaces
overlapped very well.

\section{Summary of the uniaxial case}
\label{summary}

Based on the data shown earlier in this Article for the shape and
volume changes of the void as well as the stress-strain behavior and
the stress-triaxiality, it is evident that the uniaxial loading raises
many interesting aspects to be studied in more detail. Therefore we
now concentrate on the uniaxial case when summarizing how the void
evolves and how the evolution is related to the stress-triaxiality as
the system is expanded. By plotting most of the measured values
together in one figure for the uniaxial simulation at the strain-rate
$\dot{\varepsilon} = 10^8$/sec, it is possible to compare the
evolution sequence and investigate causality, see Fig.~\ref{fig11}(a).
For clarity, we have chosen not to plot many quantities in the figure,
{\em e.g.}  plastic strain, plastic work, and temperature.  However,
their concomitant behaviors are included in the verbal explanation
below and shown in previous figures. The data shown in this figure are
from the restarted simulation (see the explanation near
Fig.~\ref{fig8}), as are the data in Figs.~\ref{fig8}-\ref{fig10}.
The mean stress and stress-triaxiality data are from that simulation,
too, and thus are different from the data shown in Figs.~\ref{fig1}(a)
and ~\ref{fig3}(a).  In any case, the overall behavior stays the same
as well as the other details as the system size, {\it etc.}

The evolution of the void and the system's stress-strain behavior can
be divided in three or even four different regions. The first region
is when the system expands elastically.  The mean and von Mises stresses
increase smoothly, nearly linearly, and the stress-triaxiality stays
nearly constant. Through the elastic region the void volume fraction
remains nearly constant, too. It is not exactly constant, since due to
the free surface of the void, the elastic expansion is a bit greater
at the surface of the void compared to the total system. Trivially the
mean and equivalent plastic strains as well as the plastic work are
equal to zero in the elastic region, and temperature decreases in the
system. The quadrupole moment has a non-zero value because of the
elongation in the direction of the strain.

The second region begins at what we call the yield point, {\it i.e.},
the onset of rapid growth of the void facilitated by plastic
deformation.  Heterogeneous nucleation of dislocations at the void
surface is the primary mechanism for plasticity in the simulation, and
thus it is at this point that the measured quantities start to deviate
from their elastic behavior. The mean stress begins to plateau here,
but fluctuating somewhat. The early departure from elastic behavior
prior to the plateau is much less pronounced. The change in the void
shape begins just at the point the mean stress deviates from elastic
behavior: $Q_{20}$ goes rapidly from the positive value acquired
during elastic expansion to a negative value, {\it i.e.}, from a
prolate shape (elongated in the direction of the strain) to an oblate
shape (expanded in the transverse directions). $Q_2$, on the other
hand, drops from a positive value, almost reaching zero at the
prolate-to-oblate transition point ($\varepsilon = 5.45$\%) and rising
even larger value after that [in fact, the oblate shape is somewhat
more pronounced than the earlier prolate shape, seen as a larger
absolute value of $Q_{20}$ in Fig.~\ref{fig10}(a)].  When $Q_2$ starts
to drop $Q_4$ starts to rise.  Then after the prolate-to-oblate
transition point, $Q_4$ begins to saturate.  At a strain of
$\varepsilon =$ 5.55\%, $Q_2$ is 1.5 times as large as its value at
the end of the elastic phase ($\varepsilon =$ 5.25\%). Mean plastic
strain, equivalent plastic strain, plastic work, and temperature
increase together with porosity. Their increase starts immediately at
the yield point, {\it i.e.}, when the mean stress first deviates from
the elastic behavior.  A bit later than the plastic strain, equivalent
plastic strain, plastic work, and temperature, the stress triaxiality
increases simultaneously with the first substantial drop in the von Mises
stress. The fact that the increase in stress-triaxiality follows later
is dependent on how the ratio between mean stress and von Mises stress
develops, as discussed earlier. In Fig.~\ref{fig3}(a) and in the data
reported in Table~\ref{tab1} the stress-triaxiality started to
increase simultaneously with the mean and von Mises stresses deviating
from the elastic behavior.  The increase of stress-triaxiality is
caused by von Mises stress plummeting in contrast to nearly constant mean
stress.  The drop in von Mises stress follows from the flow of
dislocations nucleated at the void and from the relaxation of the
shear stress of the system due to the flow.

The third region is when the void fraction, mean plastic strain,
equivalent plastic strain, plastic work, and temperature switch from
rapid increase to a linear growth or even saturate.  Subsequently the
increase of the stress-triaxiality slows down and plateaus.  The value
at the plateau is related in continuum models to the ratio of the mean
stress threshold for void growth to the flow stress.  At the plateau
von Mises stress saturates at a small value, close to the tensile
strength, and the shape of the void starts to become more of
octahedral shape although having a non-zero quadrupole moment, too.
Hence at the second and third regions the mean stress is nearly
constant, but von Mises stress and the stress-triaxiality changes.

A conclusion might be that once the threshold for void growth is
reached, the population of dislocations rises sufficiently to relax
the shear stress quite effectively and it drops to a low value (the
flow stress); the mean stress, on the other hand, plateaus since it is
relaxed by void growth and requires that the stress at the void
surface be sufficiently high to continue to nucleate dislocations.
The fourth region is the failure, when the system breaks, and it is
not studied here.

In order to see if the rapid changes studied above are due to the
smallness of the size of the void we have done one additional
simulation with a system in which the initial void radius is half that
in the other simulations; otherwise, the system size is the same, see
Fig.~\ref{fig11}(b).  For the small void simulation, uniaxial strain
at a strain-rate $\dot{\varepsilon} = 10^8$/sec has been used. Here
the difference is that the quadrupole moment suffers stronger
fluctuation due to the smallness of the void, where each of the
surface atom contribute more to its value and therefore is even more
sensitive to the selection criterion of surface atoms, and also the
shape changes are harder to determine based on the $Q_{20}$ behavior.
The other main difference is that the growth of the void is linear all
the time. Also the changes in stress-triaxiality in fact advances the
porosity when saturating, and the mean stress does not fluctuate but
drops more rapidly (this can be compared with the case without the
void, where the mean stress drops abruptly).

We have also compared the mean plastic strain and the void volume
fraction calculations. In continuum solid mechanics, it is assumed
that solid materials are plastically incompressible.  Any local
dilation, as indicated by a change in the mean strain, is attributed
either to elastic dilation or to a change in the porosity of the
material, where the porosity is equated to the void volume fraction.
The porosity $f$ and the mean plastic strain $\varepsilon^P_m$ are
then related according to the equation~\cite{Tvergaard2}
\begin{equation}
\dot{f} = 3 (1-f) \, \dot{\varepsilon}^P_m
\label{eq_delpor}
\end{equation}
where the dots denote time derivatives. Integration with respect to
time, porosity from $f_0$ to $f$, and mean plastic strain from zero to
$\varepsilon^P_m$, gives
\begin{equation}
f(\varepsilon^P_m) = 1 + (f_0 - 1 )\exp \left(-3 \varepsilon^P_m \right).
\label{eq_porosity}
\end{equation}
where $f_0$ is the initial porosity.  It is interesting to check
whether this relationship holds for the MD simulation, where other
effects such as excess volume associated with dislocation cores,
vacancies or other defects could require corrections.  In comparing
the porosity inferred from the mean plastic strain and that calculated
directly as the void fraction, the agreement is very good.  The trends
are in excellent agreement, but there is a small discrepancy between
the curves, so that the porosity from the mean plastic strain is
overestimated.  We believe that the discrepancy arises because of the
void surface.  In calculating the void fraction, we have defined the
void surface to pass through the center of the surface atoms.
However, the properties of the surface atoms are distinct from the
bulk atoms.  Therefore, there is some ambiguity in where the surface
should be placed, and a small uniform shift $\delta r$ of the surface
radially into the bulk is enough to account for the discrepancy.  In
Fig.~\ref{fig12} we have plotted the comparison of the porosity from
the mean plastic strain, Eq.~(\ref{eq_porosity}), and from the void
fraction, Eq.~(\ref{eq_void_frac}), using a constant radius increase
$\delta r = 0.58 \, a_0$, where $a_0$ is the lattice constant for the
void volume calculation.  The correction for the void size, $\delta
r$, is a fit parameter and it varies for different strain-rates and 
slightly for different
loading modes, but is always positive and of the order of the lattice
constant, $a_0$.  It should be noted, too, that by Taylor expanding
Eq.~(\ref{eq_porosity}) and discarding the higher order terms, it
becomes
\begin{equation}
f(\varepsilon^P_m) = f_0 + 3 (1-f_0) \varepsilon^P_m,
\label{eq_porosity2}
\end{equation}
indicating a linear correspondence between $f$ and $\varepsilon^P_m$
as long as the void fraction is small.

\section{Conclusions and Discussions}
\label{discussion}
In this Article void growth in copper has been studied in a high range
of strain-rates at the atomistic level.  The model has been designed
to simulate the growth of a void nucleating from a very weakly bound
inclusion during strain-controlled dynamic fracture.  In order to see
the effect of various modes of expansion and the related
stress-triaxiality, three different modes have been applied, namely
uniaxial, biaxial, and triaxial. The molecular dynamics method
developed here has been shown to be efficient enough to explore the
different loading conditions and strain-rates varying over four orders
of magnitude. A uniform expansive loading of a system with periodic
boundary conditions has been implemented using a well defined scaling
matrix method. For the longest calculations, the MD method was
parallelized successfully. The macroscopic observables mean stress,
von Mises stress, stress-triaxiality, mean plastic strain, equivalent
plastic strain, plastic work, and temperature have been calculated and
compared with the microscopic quantities measured at the
atomistic-level, such as the volume of the void and its shape
change. A method to describe the shape changes in the void is
introduced and employed, namely calculations of the multipole moments
of the void based on spherical harmonics in polynomial, not
trigonometric, form. When calculating the volume of the void with an
unknown shape or defining solid surface for the multipole moment
calculation a useful method, namely optimal triangular tessellation,
has been introduced.  This method has been extended from the usual
planar case to non-planar objects such as the surface of the void.

When the different measured quantities are compared with each other
during an MD simulation in uniaxial expansion, it is found that at
early stages of plasticity von Mises stress, and thus also
stress-triaxiality, plays a more significant role to the void growth
and its shape change than expected. On the other hand, most of the
macroscopic plastic quantities as mean and equivalent plastic strain
as well as plastic work and temperature, seem to be more dependent on
the simultaneous saturation of the mean stress. These calculations
show a counter-intuitive behavior, observed previously in continuum
void growth modeling,~\cite{castaneda,budiansky,Andersson} that a
prolate-to-oblate transition occurs.  When the system starts to yield,
the expansion of the void switches from its original elastic extension
in the direction of the load to transverse plastic expansion.

The yield stress values for the lowest strain-rates $10^7$/sec are in
reasonable agreement with the experimentally measured spall
strength.~\cite{Kanel} The fact that mean plastic strain can be mapped
to the growth of the void is consistent with continuum
models.~\cite{Tvergaard2}

This study leaves many open questions. For instance related to the
void growth are the dislocations, which occur when the system
yields. Since the FCC crystal studied here is perfect apart from the
void, the dislocations form from void's surface. They are also
responsible for its growth by carrying material away. Thus the
characterization of plasticity surrounding a growing void at the level
of dislocations should be investigated, too, especially with respect
to the stress-triaxiality. These investigations are
underway.~\cite{unpubl} Their results are beyond the scope of this
Article, other than to mention that the identification of the yield
point in this Article does indeed correspond to the point of initial
nucleation of dislocations.  Another topic that is beyond the scope of
this Article and needs further investigation, but is closely related
to the studies here, is the quantitative connection between the shear
stress, and thus the mode of the loading, and the onset of the void
growth and the resulting change in the stress state.  Other areas
where this study can be extended are different materials including
different lattice structures such as body-centered cubic (BCC)
lattices; in the uniaxial case other orientations of the lattice as
$\langle 110 \rangle$ and $\langle 111 \rangle$; continuously changing
stress-triaxiality in order to create the full yield surface to the
stress space similarly as in Gurson type of continuum
studies;~\cite{Gurson} to include grain boundaries, defects,
pre-existing dislocations, several voids, etc.

\begin{acknowledgments}

This work was performed under the auspices of the U.S.\ Dept.\ of
Energy by the University of California, Lawrence Livermore National
Laboratory, under contract number W-7405-Eng-48. The authors would
like to thank Dr.\ Richard Becker bringing to our attention
Refs.~\onlinecite{budiansky} and~\onlinecite{Andersson} that find the
prolate-oblate transition in continuum modeling.

\end{acknowledgments}

\appendix

\section{Multipole moments}
\label{app1}

The 24 different polynomial terms of the multipole moments used in this
study are listed below.~\cite{Stone} Here the conventional notation is 
used, so that the principal axis of the coordinates is the $z$-direction.  
When these formulas are used in interpreting the void shape evolution, 
the principal axis is the direction with uniaxial loading, {\it i.e.}, 
the $x$-coordinate in the Article. Similarly the load free directions 
$y$ and $z$ correspond to $x$ and $y$ below.

The polynomial terms when $l=1$ are the dipole moments and they
capture if the object is offset. The dipole moments are:
\begin{equation}\begin{array}{l}
\mbox{\hspace{5mm}} Q_{10} = \frac{1}{2} \sqrt{ \frac{3}{\pi}} 
\frac{1}{\bar{r}^2}  \int \! r z \, d\Omega, \\
\mbox{Re } Q_{11} = - \frac{1}{2} \sqrt{ \frac{3}{2\pi}} \frac{1}{\bar{r}^2} 
\int \! r x \, d\Omega, \\
\mbox{Im } Q_{11} = - \frac{1}{2} \sqrt{ \frac{3}{2\pi}} \frac{1}{\bar{r}^2} 
\int \! r y \, d\Omega,\\
\label{eq_ql1}
\end{array}
\end{equation}
where $\bar{r}^2 = \frac{1} {4 \pi} \int r^2(\theta,\phi) d\Omega$.

Terms with $l=2$ are the quadrupole moments and they get non-zero
values if there is ellipsoidal shape in the object. The quadrupole
moments are as follows:
\begin{equation}\begin{array}{l}
\mbox{\hspace{5mm}} Q_{20} = \frac{1}{4} \sqrt{ \frac{5}{\pi}} 
\frac{1}{\bar{r}^2} \int  \! 3 z^2 - r^2 \, d\Omega, \\
\mbox{Re } Q_{21} = -\frac{1}{2} \sqrt{ \frac{15}{2\pi}} \frac{1}{\bar{r}^2} 
\int \!  xz \, d\Omega, \\
\mbox{Im } Q_{21} = -\frac{1}{2} \sqrt{ \frac{15}{2\pi}} \frac{1}{\bar{r}^2}
\int \! yz \, d\Omega,\\
\mbox{Re } Q_{22} = \frac{1}{4} \sqrt{ \frac{15}{2\pi}} \frac{1}{\bar{r}^2}
\int \! x^2 - y^2 \, d\Omega,\\
\mbox{Im } Q_{22} = \frac{1}{2} \sqrt{ \frac{15}{2\pi}} \frac{1}{\bar{r}^2}
\int \! xy \, d\Omega.
\label{eq_ql2}
\end{array}
\end{equation}

Terms with $l=3$ are the octupole moments and they get non-zero values
for tetrahedron shapes:
\begin{equation}\begin{array}{l}
\mbox{\hspace{5mm}} Q_{30} = \frac{1}{4} \sqrt{ \frac{7}{\pi}} 
\frac{1}{\bar{r}^2} \int \! \frac{1}{r} z ( 5 z^2 -3 r^2) \, d\Omega, \\
\mbox{Re } Q_{31} = - \frac{1}{8} \sqrt{ \frac{21}{\pi}} \frac{1}{\bar{r}^2}
\int \! \frac{1}{r} x (5z^2  - r^2) \, d\Omega, \\
\mbox{Im } Q_{31} = -\frac{1}{8} \sqrt{ \frac{21}{\pi}} \frac{1}{\bar{r}^2}
\int \! \frac{1}{r} y (5z^2 - r^2)  \, d\Omega,\\
\mbox{Re } Q_{32} = \frac{1}{4} \sqrt{ \frac{105}{2\pi}} \frac{1}{\bar{r}^2} 
\int \! \frac{1}{r} z (x^2  -y^2) \, d\Omega,\\
\mbox{Im } Q_{32} = \frac{1}{2} \sqrt{ \frac{105}{2\pi}} \frac{1}{\bar{r}^2}
\int  \! \frac{1}{r} xyz \, d\Omega,\\
\mbox{Re } Q_{33} = -\frac{1}{8} \sqrt{ \frac{35}{\pi}} \frac{1}{\bar{r}^2}
\int \! \frac{1}{r} (x^3- 3xy^2) \, d\Omega,\\
\mbox{Im } Q_{33} = \frac{1}{8} \sqrt{ \frac{35}{\pi}} \frac{1}{\bar{r}^2} 
\int \!\frac{1}{r}  (y^3- 3x^2y) \, d\Omega. 
\label{eq_ql3}
\end{array}
\end{equation}

And finally the terms with $l=4$ are listed. They are the 
hexadecapole moments and capture octahedron shapes:
\begin{equation}\begin{array}{l}
\mbox{\hspace{5mm}} Q_{40} = \frac{3}{4} \sqrt{ \frac{1}{\pi}} 
\frac{1}{\bar{r}^2} \int \! \frac{1}{r^2} ( 3r^4 -30 r^2 z^2 + 35 z^4) \, d\Omega, \\
\mbox{Re } Q_{41} = - \frac{3}{8} \sqrt{ \frac{5}{\pi}} \frac{1}{\bar{r}^2}
\int \! \frac{1}{r^2} xz (7z^2  - 3r^2)  \, d\Omega, \\
\mbox{Im } Q_{41} = - \frac{3}{8} \sqrt{ \frac{5}{\pi}} \frac{1}{\bar{r}^2}
\int \! \frac{1}{r^2} yz (7z^2  - 3r^2)  \, d\Omega, \\
\mbox{Re } Q_{42} = \frac{3}{8} \sqrt{ \frac{5}{2\pi}} \frac{1}{\bar{r}^2}
\int \! \frac{1}{r^2} (x^2  -y^2)(7z^2 -r^2)  \, d\Omega,\\
\mbox{Im } Q_{42} = \frac{3}{4} \sqrt{ \frac{5}{2\pi}} \frac{1}{\bar{r}^2}
\int \! \frac{1}{r^2} xy (7z^2 - r^2)  \, d\Omega,\\
\mbox{Re } Q_{43} = -\frac{3}{8} \sqrt{ \frac{35}{\pi}} \frac{1}{\bar{r}^2}
\int \! \frac{1}{r^2} (x^3- 3xy^2)z  \, d\Omega,\\
\mbox{Im } Q_{43} = -\frac{3}{8} \sqrt{ \frac{35}{\pi}} \frac{1}{\bar{r}^2}
\int \! \frac{1}{r^2} (3x^2y - y^3) z  \, d\Omega.\\
\mbox{Re } Q_{44} = \frac{3}{16} \sqrt{ \frac{35}{2 \pi}} \frac{1}{\bar{r}^2}
\int \! \frac{1}{r^2} (x^4- 6x^2y^2 + y^4 )  \, d\Omega,\\
\mbox{Im } Q_{44} = \frac{3}{4} \sqrt{ \frac{35}{2 \pi}} \frac{1}{\bar{r}^2}
\int \! \frac{1}{r^2} xy (x^2- y^2)  \, d\Omega.
\label{eq_ql4}
\end{array}
\end{equation}


\begin{figure}
\caption{Mean stress $\sigma_m$ versus engineering strain
$\varepsilon$ for strain-rates $\dot{\varepsilon} = 10^{10}$/sec, 
$10^9$/sec, $5 \times 10^8$/sec, $10^8$/sec, and $10^7$/sec.  
The equilibrium size of the
simulation box is $[21.7 \mbox{nm}]^3$, when $\sigma_m = 0$. The
simulation box has 860~396 atoms and a pre-existing void of radius 2.2
nm. The thin solid line, drawn as a reference, is from a system with
no initial void, consisting of 364~500 atoms in an equilibrium box
sized $[17.5 \mbox{nm}]^3$, and expanded at $\dot{\varepsilon} =
10^9$/sec.  (a) Uniaxial expansion with $\varepsilon_x = \varepsilon$,
$\varepsilon_y = \varepsilon_z = 0$. The inset zooms on the yield
points of the stress-strain curve. (b) Biaxial expansion with
$\varepsilon_y = \varepsilon_z = \varepsilon$, $\varepsilon_x =
0$. (c) Triaxial expansion with $\varepsilon_x = \varepsilon_y =
\varepsilon_z = \varepsilon$.}
\label{fig1}
\end{figure}

\begin{figure}
\caption{von Mises stress $\sigma_e$ versus engineering strain
$\varepsilon$ from the same simulations as in Fig.~\ref{fig1}.  In
uniaxial and biaxial expansion, von Mises stress rises until the onset
of void growth and then it drops to a small value; in triaxial
expansion it is always small.  See the caption of Fig.~\ref{fig1} for
simulation details.  (a) Uniaxial, (b) biaxial, and (c) triaxial
expansion. }
\label{fig2}
\end{figure}

\begin{figure}
\caption{The stress-triaxiality $\chi$ (\ref{eq_triax}) versus
engineering strain $\varepsilon$ from the same simulations as in
Fig.~\ref{fig1}.  See the caption of Fig.~\ref{fig1} for the details.
(a) Uniaxial and (b) biaxial expansion. In triaxial expansion
stress-triaxiality is diverging and not defined.}
\label{fig3}
\end{figure}

\begin{figure}
\caption{Mean plastic strain $\varepsilon_m^P$, calculated using
Eq.~(\ref{eq_mean_plastic}), versus engineering strain $\varepsilon$
from the same simulations as in Fig.~\ref{fig1}.  See the caption of
Fig.~\ref{fig1} for the details.  (a) Uniaxial, (b) biaxial and (c)
triaxial expansion.}
\label{fig4}
\end{figure}

\begin{figure}[ht]
\caption{Equivalent plastic strain $\varepsilon_e^P$, calculated as
Eq.~(\ref{eq_equivalent2}), versus engineering strain $\varepsilon$
from the same simulations as in Fig.~\ref{fig1}.  See the caption of
Fig.~\ref{fig1} for the details.  (a) Uniaxial and (b) biaxial
expansion.}
\label{fig5}
\end{figure}

\begin{figure}[ht]
\caption{Plastic work $W_P$, calculated from
Eq.~(\ref{eq_plastic_work}), versus engineering strain $\varepsilon$
from the same simulations as in Fig.~\ref{fig1}.  See the caption of
Fig.~\ref{fig1} for the details.  (a) Uniaxial, (b) biaxial, and (c)
triaxial expansion.}
\label{fig6}
\end{figure}

\begin{figure}
\caption{Temperature $T$, versus engineering strain $\varepsilon$ from
the same simulations as in Fig.~\ref{fig1}.  Compare with the plastic
work plotted in Fig.~\ref{fig6}.  See the caption of Fig.~\ref{fig1}
for the details.  (a) Uniaxial, (b) biaxial, and (c) triaxial
expansion.  }
\label{fig7}
\end{figure}

\begin{figure}[ht]
\caption{Void volume fraction versus engineering strain $\varepsilon$.
The evolution of the ratio of the void volume to the total box volume
is plotted for strain-rates $\dot{\varepsilon} = 10^{10}$/sec,
$10^9$/sec, $5 \times 10^8$/sec, and $10^8$/sec. See the caption of
Fig.~\ref{fig1} for additional details.  (a) Uniaxial, (b) biaxial,
and (c) triaxial expansion.}
\label{fig8}
\end{figure}

\begin{figure}
\caption{Snapshots of the atoms comprising the surface of the void
during uniaxial expansion with $\varepsilon_x = \varepsilon$,
$\varepsilon_y = \varepsilon_z = 0$.  The simulation box is oriented
along $\langle 100 \rangle$ directions, so that the $z$-axis is out of
the paper. The strain-rate is $\dot{\varepsilon} = 10^8$/sec. See the
caption of Fig.~\ref{fig1}(a) for additional details.  The panels show
snapshots at different strains: (a) $\varepsilon = 5.05\%$ (b)
$\varepsilon = 5.26\%$ (c) $\varepsilon = 5.47\%$ (d) $\varepsilon =
5.68\%$ (e) $\varepsilon = 5.89\%$ (f) $\varepsilon = 6.10\%$}
\label{fig9}
\end{figure}

\begin{figure}
\caption{Multipole moments of the void surface calculated using
Eq.~(\ref{eq_qlm}).  (a) Quadrupole moment $Q_{2m}$ with $m=0,1,2$ for
uniaxial expansion at $\dot{\varepsilon} = 10^8$/sec. 
(b)-(d) The moments $Q_{l}$ for $l=1,2,3,4$ in (b)
uniaxial, (c) biaxial, and (d) triaxial cases. They are calculated
using Eq.~(\ref{eq_ql}) and the strain-rate $\dot{\varepsilon} =
10^8$/sec.  See the caption of Fig.~\ref{fig1} for details of the
simulations.}
\label{fig10}
\end{figure}

\begin{figure}
\caption{(a) The mean stress $\sigma_m$ (thick solid line),
stress-triaxiality $\chi$ (dotted line), volume fraction $f$ of the
void (dashed line), and the quadrupole moment $Q_{20}$ (thin solid
line) from the simulation with uniaxial expansion at
$\dot{\varepsilon}=10^8$/sec.  See the captions of
Figs.~\ref{fig1},~\ref{fig3},~\ref{fig8}, and~\ref{fig10} for the
details. (b) As a comparison the same measures as in (a), but now for
the case having an initial void radius of 1.1~nm and 863~543 atoms in
the system undergoing uniaxial expansion at the same
$\dot{\varepsilon}=10^8$/sec.}
\label{fig11}
\end{figure}

\begin{figure}
\caption{Porosity $f$ calculated from the actual void fraction as in
Eq.~(\ref{eq_void_frac}) with $\delta r = 0.58 \, a_0$ (see the text
for details of $\delta r$) and from the mean plastic strain
$\varepsilon_m^P$ as in Eq.~(\ref{eq_porosity}) from the simulation
with uniaxial expansion at $\dot{\varepsilon}=10^8$/sec.  See the
captions of Figs.~\ref{fig1},~\ref{fig4},~\ref{fig8}, and~\ref{fig11}
for the simulation details.}
\label{fig12}
\end{figure}

\end{document}